# Using quantum transport model to probe molecule-induced tuning of semiconductor band-bending

*Archana Bahuguna, Fernanda Camacho-Alanis, Riya Shergill, Smitha Vasudevan, Avik W. Ghosh, Nathan Swami**

Department of Electrical and Computer Engineering, University of Virginia,

Charlottesville, Virginia 22904 (USA)

nswami@virginia.edu

ABSTRACT Electron transfer processes at molecule-semiconductor interfaces involve a complex mixture of thermionic, tunneling and hopping events. Traditionally these processes have been modeled in a piece-meal fashion, relying on phenomenological treatments such as Simmons and Richardson equations that are not vetted in atomistic systems and do not flow seamlessly into each other. We present a unified modeling approach, based on the Non-equilibrium Green's function (NEGF) formalism that allows us to integrate diverse transport regimes and establish a comprehensive quantitative theory. By comparing our simulations with experiments on a metal-molecule-semiconductor junction (varying molecular lengths ~1-3nm), we identify the role of molecular monolayers in tuning the semiconductor band-bending, and thereby overall device conductivity. We find that the principal role of molecules is to act as a voltage divider, altering the Schottky barriers, thereby modulating current levels, voltage-asymmetries and crossover from Schottky to tunneling transport. While this provides an appealingly simple explanation for our observed experimental trends, the calculated shifts in crossover voltages are insufficient to explain the experiments quantitatively. Quantitative correspondence with experiments requires invocation of an additional voltage divider arising from molecular dipoles that further tunes semiconductor band-bending. The extracted dipole moments are rationalized using *ab-initio* calculations for each molecule, along-with a dilution of packing fraction for the shortest molecular lengths. The methodology described herein can be used to better understand and predict transport characteristics of such junctions.

Organic molecules can critically influence the performance of electronic devices such as transistors and solar cells through their functionally tunable transmissions, surface charge modulation, guided self-assembly, and the generation of molecule-specific transport signatures. These signatures include non-linear variations in conductivity such as switching, memory and negative differential resistance [1], as well as shape changes in the current density-voltage (J-V) characteristics of rectifying junctions composed of insulating molecules due to molecule-induced variations to the semiconductor Schottky barrier. Charges and dipoles at the semiconductor surface, which arise due to bonding with the molecule, can influence semiconductor band-bending and hence, the Schottky barrier. The band-bending can thereby modulate the semiconductor channel conductivity between Source and Drain contacts in field-effect transistors (FET) or passivate surface states within solar cells. Tuning the band-bending will require characterization techniques sensitive to variations in interfacial charge configurations. Traditionally we have relied on photoemission spectroscopy to extract information on the density and energy distribution of interface states, while work function measurements probed changes in electron affinity at the semiconducting surface upon molecular adsorption. [2] However, these methods present only averaged information over large scaled features. In contrast, electrical transport measurements are highly sensitive to detection of local electro-active defects and can be used to probe sub-micron scale features. [3] To enable molecule-induced tuning of semiconductor band-bending, there is thus a need to develop accurate methods that can couple J-V characteristics with quantitative transport models to probe the interfacial barriers.

Traditionally one calculates J-V characteristics using phenomenological approaches such as the Simmons model for symmetric junctions [4] or the Brinkman model for asymmetric junctions [5]. These models usually fit select voltage ranges of the J-V characteristics to determine parameters such as barrier height, width, and effective mass values that characterize transport in the voltage regime. For instance, the Simmons model has been widely applied over the last decade to compare molecular junction conductance on Au, [6] GaAs [7] and Si. [8] However, these phenomenological models cannot quantitatively de-convolve the role of individual components such as interfacial charges, molecular packing density, induced-dipoles, and Fermi-level pinning of interface states by molecule [9]. Furthermore, the models were developed for bulk, isotropic systems and are not vetted for typically anisotropic, lower-dimensional structures. For the purpose of understanding and guiding the molecular tuning of device interfaces, there is a pressing need for a rigorous theoretical model, such as the Non-Equilibrium Green's Function (NEGF) [10] formalism of quantum transport.

In this paper we aim to couple the experimental J-V characteristics [11] (Figure 1) of MMS junctions to a quantum transport model for quantitative determination of barriers resulting from molecule-induced interfacial charges at the semiconductor surface. As reported within prior work on GaAs [7, 11, 12] and Si, [13] the experimental J-V characteristics depict distinct slopes in reverse and forward bias regimes. As shown in Figure 1, we observe a prominent kink in the forward bias J-V that arises from a cross-over between low-bias thermionic emission over the GaAs Schottky barrier and high-bias quantum mechanical tunneling across the molecular tunnel barrier. The former is controlled by the semiconductor depletion width (doping and polarizability) while the latter depends on the molecular length and gas-phase polarizability. Increasing the molecular length reduces the overall current, simultaneously dropping much of the applied voltage across the molecule and thereby reducing the semiconductor

Schottky barrier. The result is a progressive reduction in cross-over voltage and a corresponding increase in J-V voltage symmetry. We first attempt to analyze the J-V characteristics using the traditional Richardson and Simmons approach to enable a discussion of its inadequacies. A more elaborate quantum transport model of the MMS junctions based on the NEGF formalism follows, which was applied to quantitatively determine the molecule-induced interfacial barriers that can be applied towards the tuning of semiconductor band-bending.

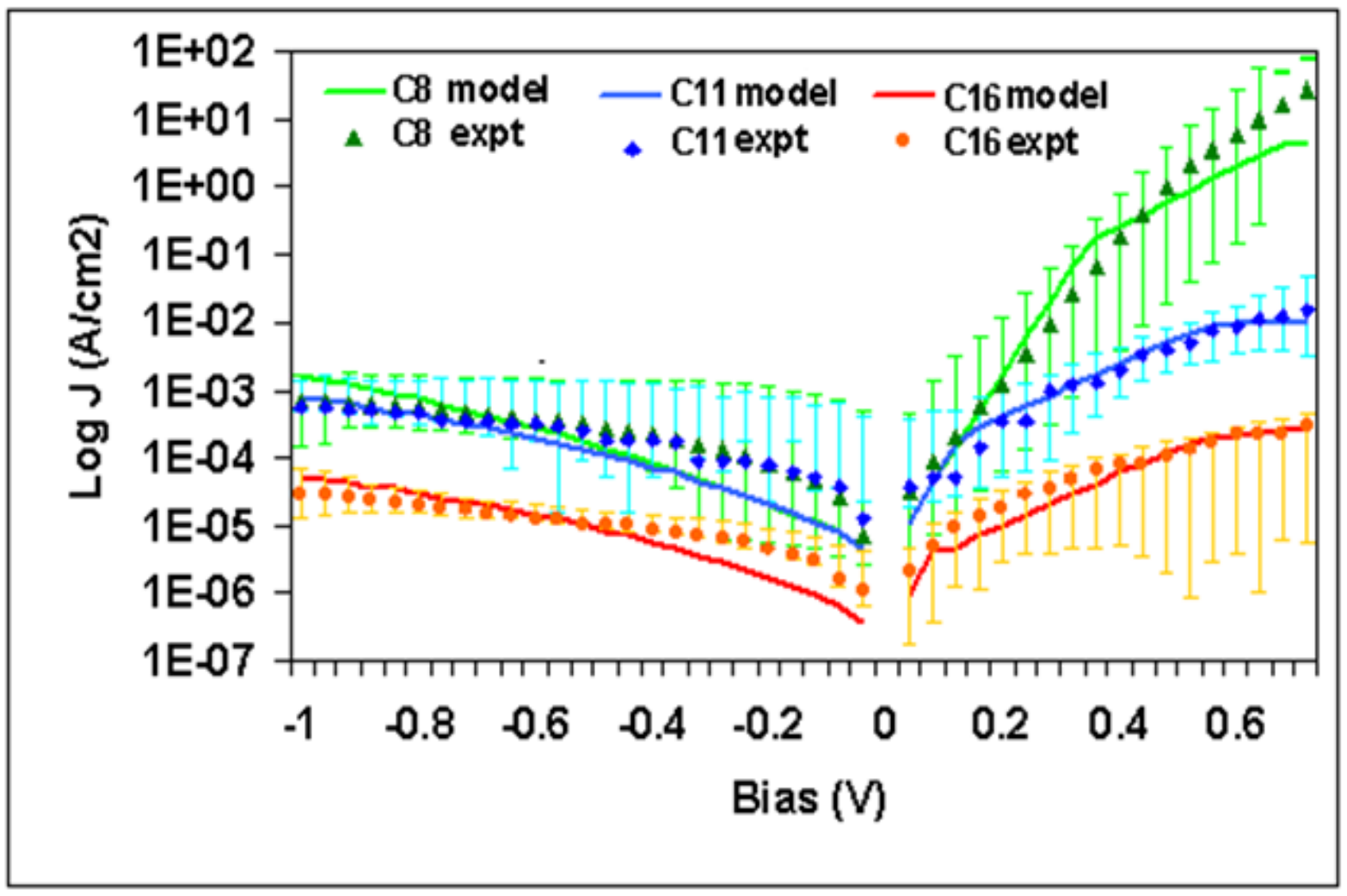

**Figure 1.** Experimental J-V characteristics (symbols with error bars for ~35 devices per molecule-type) for molecular junctions of different device lengths (C8, C11 and C16). The optimized fits based on the quantum transport model (smooth curves) are used to extract interfacial barriers, as explained in Section 5.

We set up NEGF equations coupled to a continuum model for the MMS junction, with the band-bending determined by electrostatic boundary conditions. While more elaborate 'first-principles' density functional theories are in vogue, [14, 15] their extension to semiconductor contacts bearing long depletion widths and complex surface bandstructures is considerably challenging. [16, 17] Often, equilibrium Quantum chemical calculations are performed on isolated molecules - molecular orbital levels, density of states and dipole moments to substantiate experimental results on MMS junctions. [13] We focus instead on the physics at hand, namely, voltage division across the various junction segments, as well as off-resonant tunneling through the wide molecular band-gaps. Since the molecular levels are seen to be sitting far from the conducting window and play little active role in the transport, we choose to simplify our molecular description in lieu of a continuum model, analogous to those adopted for amorphous organic molecular layers.

The simplicity of our band-structure and the universality of our transport equations allow us to extract insights into the shapes of the observed J-V characteristics. By pushing a quantitative agreement between theory and experiment, we were able to explain how the Schottky (depletion width) and tunnel barriers (molecular length) determine the J-V shapes and magnitudes. To further fine-tune the fit to experiment (especially the "kink" voltage values and J-V symmetry) we need an additional voltage drop across a charge double layer. We attribute this layer to the creation of an interfacial dipole at the molecule-semiconductor interface due to charge transfer as well as induced dipoles (image charges). The extracted dipole moments can be rationalized with *ab-initio* calculations, assuming a reasonable packing fraction for each molecular species. Based on this we envision that this particular quantum transport model coupled to transport measurements on MMS junctions can be applied to quantitatively probe the molecule-induced tuning of semiconductor band bending, which can be adopted eventually towards the modulation of conductivity within field effect transistor (FET)-like three terminal device paradigms. [18]

## Results and discussion

Molecular junctions were formed on an n-GaAs substrate (doped $10^{18}/cm^3$) with self assembled monolayers (SAMs) of alkanethiol molecules of varying lengths ($SH-(CH_2)_n-COOH$, n = 7, 10 and 15); henceforth called C8, C11 and C16 SAMs, respectively. The terminal COOH group enabled the bonding of a high-fidelity Cu top contact to the molecular layer through complexation and electroless deposition (see Methods section for more detail).

*Richardson Simmons results*

First, we applied the Richardson's-Simmons model (see Methods) fit to the experimental J-V characteristics. While Richardson's equation[19] was used to match the Schottky-dominated reverse and low forward bias regions, Simmons equation was used for the tunneling-dominated forward bias regions of the J-Vs. Figure 2a depicts the band diagram of the device under reverse bias showing voltage drops $V_1$ and $V_2$ across the Schottky and tunnel barriers respectively. Figure 2b depicts the results for molecular junctions with C16 SAMs (other molecular layers gave similar results).

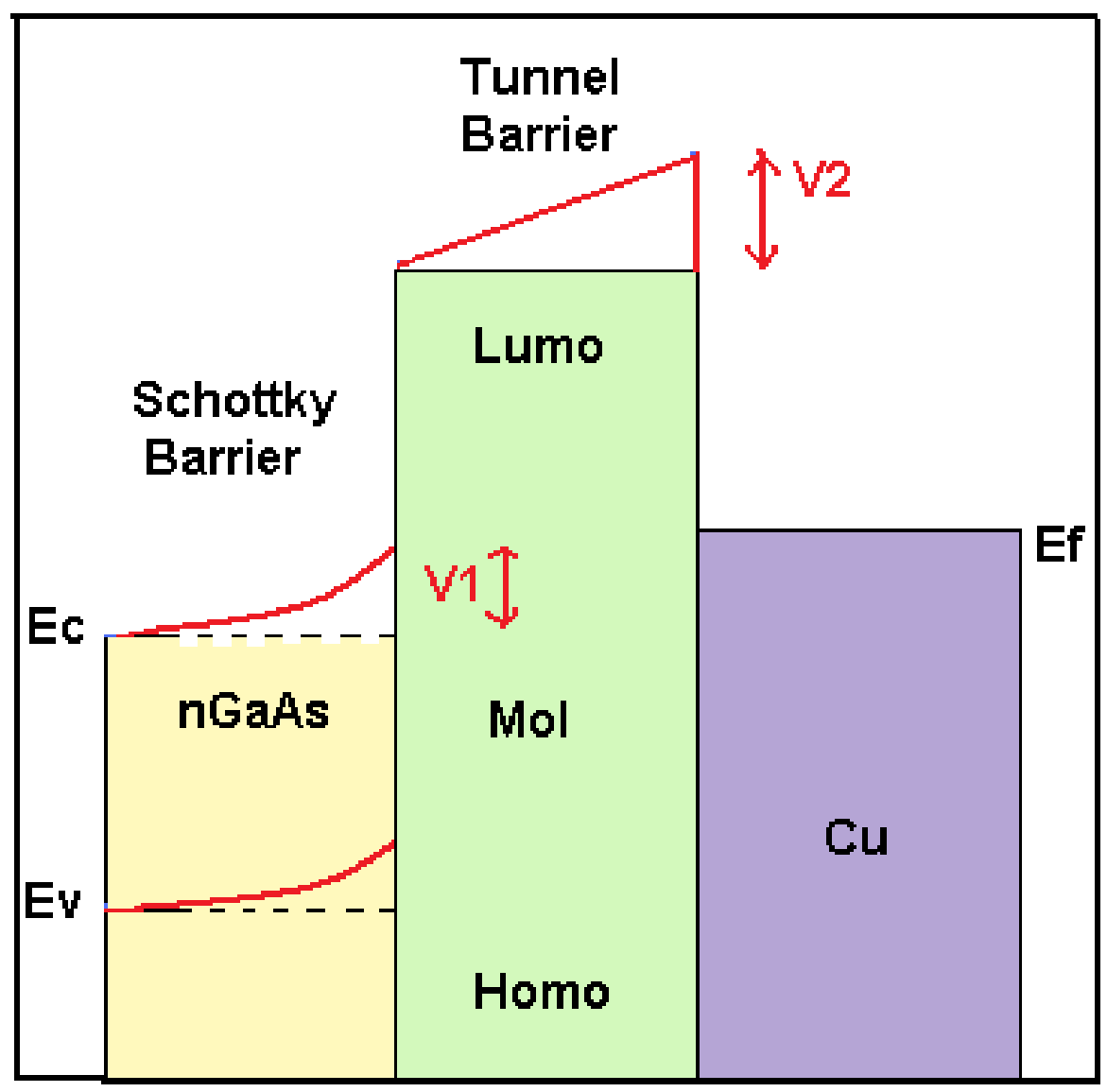

**Figure 2a.** Energy band diagram of the device under reverse bias showing the voltage drops $V_1$ and $V_2$ across the Schottky and Tunneling barriers respectively.

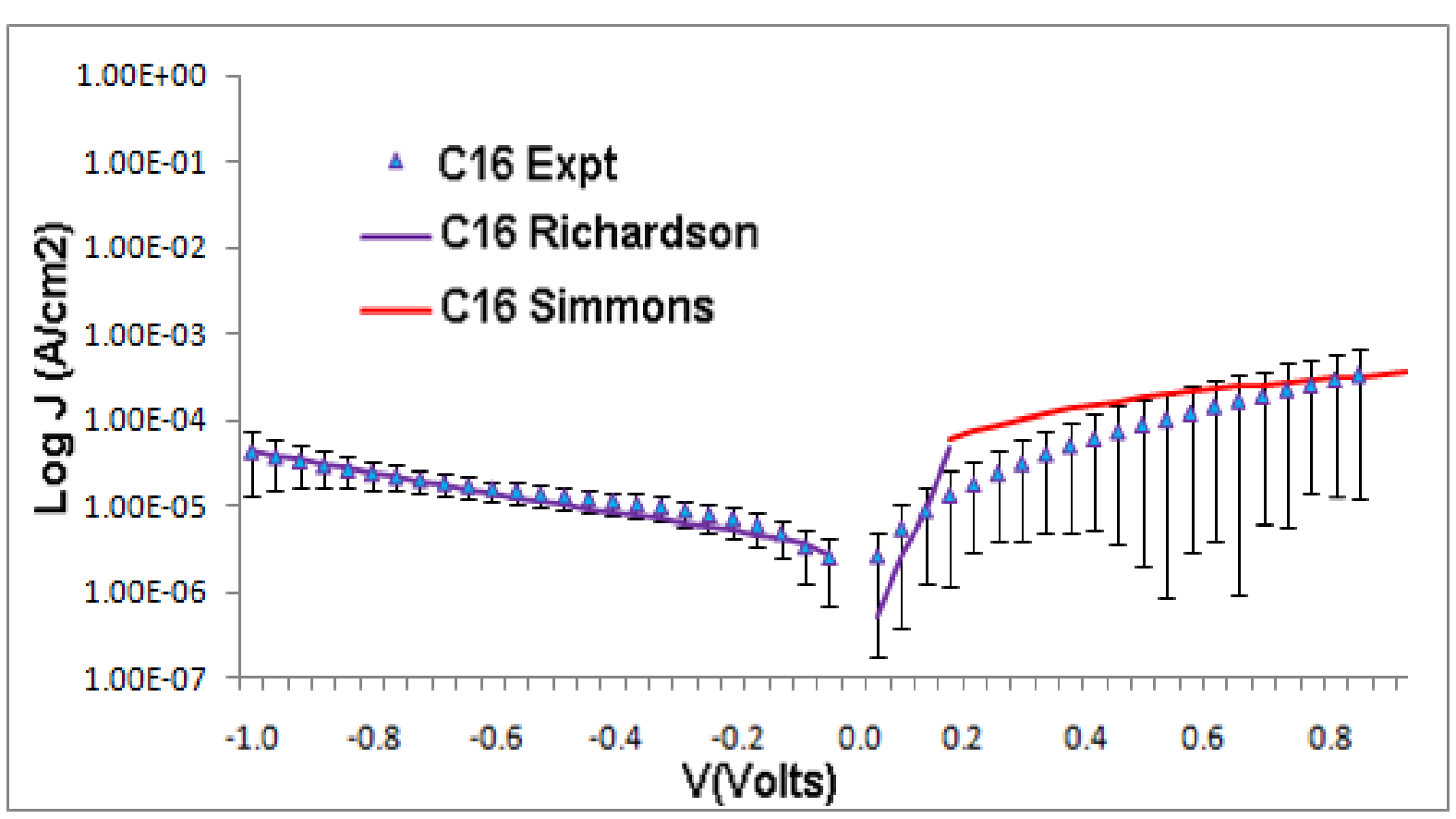

**Figure 2b.** Plots using Richardson-Simmons model for C16 compared with experiment.

In reverse bias, the Richardson equation fit matched well to experiment with respect to the current magnitudes as well as slopes. At low forward bias however, while the slopes were comparable to experiment, a quantitative comparison to the experimental current magnitudes required the invocation of a higher Schottky diode ideality factor (n~1.5, instead of ~1). This indicates that while thermionic emission alone can explain the transport mechanism through the MMS junction at reverse bias, other mechanisms such as tunneling start to become significant at low forward bias and the electron transport becomes a convolution of the two factors. At high forward bias, the fits obtained using only thermionic emission resulted in much higher slopes than experimental data and only matched experiment when ideality factors were increased to beyond n~3. This further substantiates the diminishing role of thermionic emission at successively higher forward bias values, with tunnel transport being the dominant mechanism. Hence, the Simmons equation, which has been used previously to model tunnel transport was used to fit the data for the high forward bias regions, and these match well with experimental characteristics.

Figure 3 depicts how voltage drops across the Schottky and tunneling regions, $V_1$ and $V_2$, vary with applied bias thus providing an insight on the underlying transport mechanisms through the MMS device. Considering the GaAs depletion width and the molecular layer as two electrical resistive components in the circuit, we can observe how the GaAs depletion width dependent Schottky barrier dominates at reverse and low forward biases ($V<0.2V$), as apparent from dominant variations in the respective voltage drop $V_1$ with respect to applied voltage (V) versus significantly smaller $V_2$ variations; while molecular layer dependent tunnel transport dominates at high forward bias ($V\geq0.2V$), as apparent from variations in the respective voltage drop $V_2$ versus significantly smaller $V_1$ variations. The figure also shows that this transition from Schottky to Tunneling occurs at the kink value 0.2V for C16, which is where the depletion width $W_d$ becomes equal to the Debye length ($L_d$). The barriers can thus be treated as two nonlinear resistors such that the Schottky barrier dominates at low bias leading to good fits based on Richardson analysis, while the molecular tunnel barrier dominates at high bias leading to reasonably good fits based on Simmons model analysis. The discrepancies with the Richardson-Simmons models indicate that the dominance is not absolute, and the tunneling contributions through the Schottky barrier also play a non-negligible role throughout the voltage range.

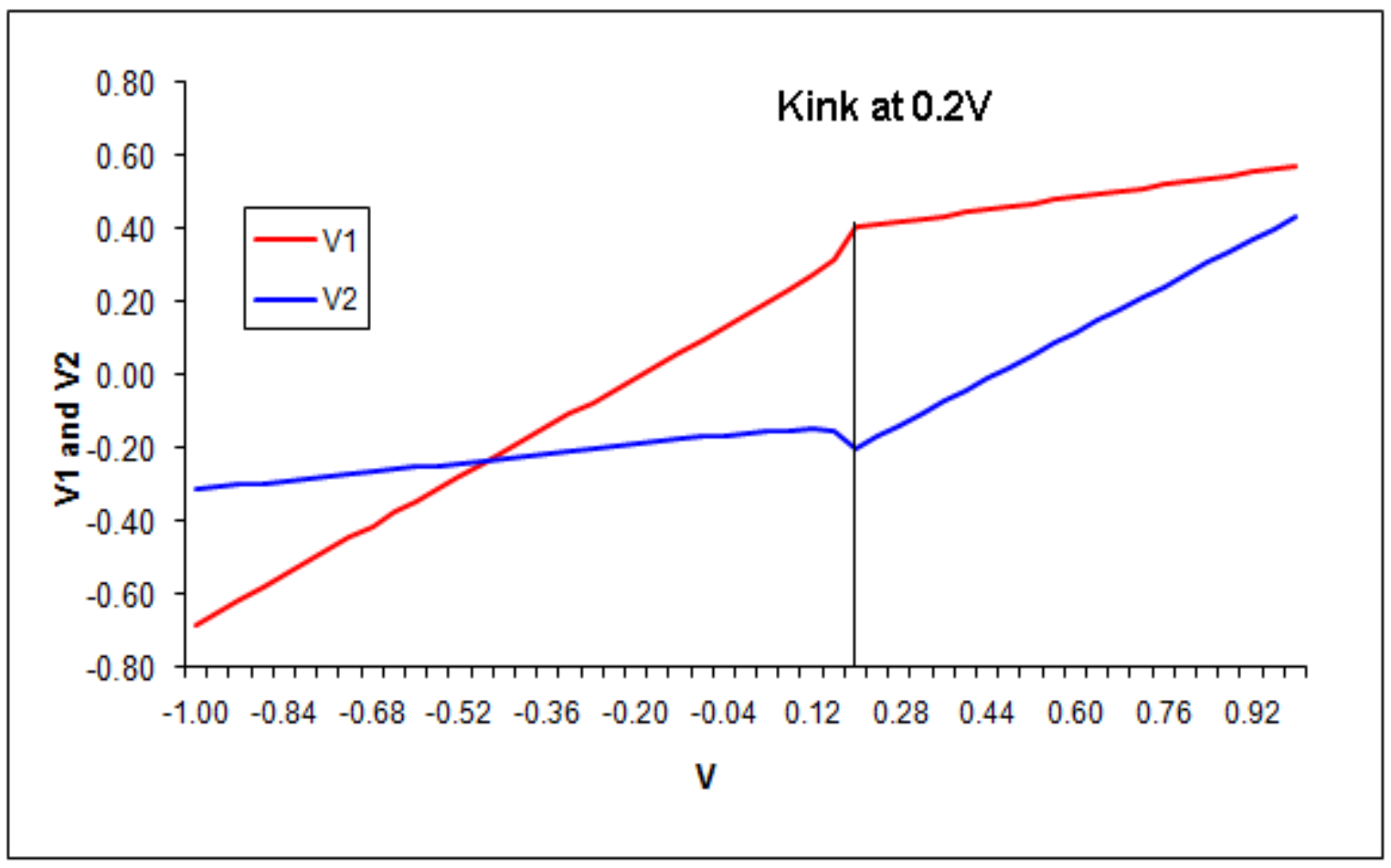

**Figure 3.** Dependence of computed voltage divisions $V_1$ and$V_2$ with applied bias for C16. Prior to the kink at V~0.2V for C16, $V_1$ varies with V while $V_2$ remains essentially constant, while after the kink $V_1$ becomes essentially constant and $V_2$ varies. Under these conditions, one can think of the heterojunction as two voltage dividing nonlinear resistors in series – the first dominates at low bias while the second dominates at high bias.

We conclude that although the Richardson and Simmons models provided us with useful insights about the experimental J-V transport characteristics, they were mostly qualitative in nature, broadly providing an overview of different transport mechanisms dominant in the different bias regimes. The outstanding issues requiring further explanation include why the kinks occur at the particular voltage values while transitioning from Schottky to Tunneling mediated transport, why the kinks shift or the J-V symmetry changes with molecular length, and understanding the effect of interface-charges/barriers/dipoles on transport. Moreover, the use of a lumped parameter such as ß, which is dependent on both the effective mass as well as the barrier for tunneling, cannot provide values of the individual parameters. Finally, these models also do not incorporate molecular properties like packing density and dielectric constant. Hence, we aim to explain these effects using our bottom-up NEGF model approach.

### *NEGF model results*

We used a continuum effective mass model for the MMS junction, keeping in mind the irrelevance of the detailed orbital chemistry of the molecular barrier in the tunneling regime (see Methods section). The molecule was treated as a simple dielectric, since its HOMO and LUMO levels sit far from the GaAs band-edges. Considering the semiconductor depletion width, the length of the molecule and a few atomic layers of the metal as the active part of the device (see grid model in Figure 4a and energy model in Figure 4b), we defined the Hamiltonian matrix (Figure 4c), H representing the electron energy profiles (near conduction band for nGaAs and LUMO level for molecule) modulated by the applied bias defined as potential matrix U. The current at an applied bias was finally calculated by solving a system of transport equations based on Green's function formalism.

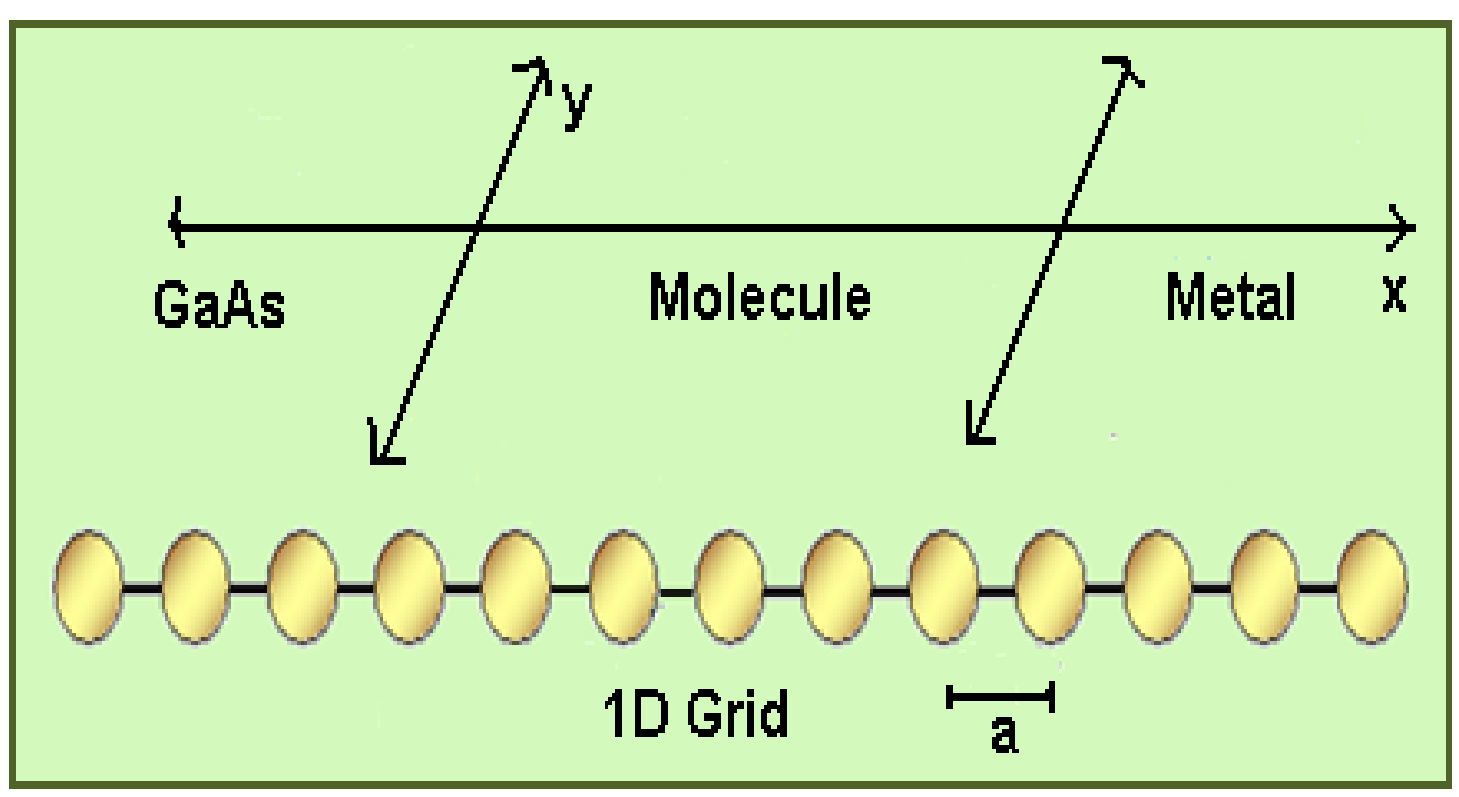

**Figure 4a.** 1D grid of the model device, with the transverse modes subsumed into the 2D Fermi functions in Eq.[7].

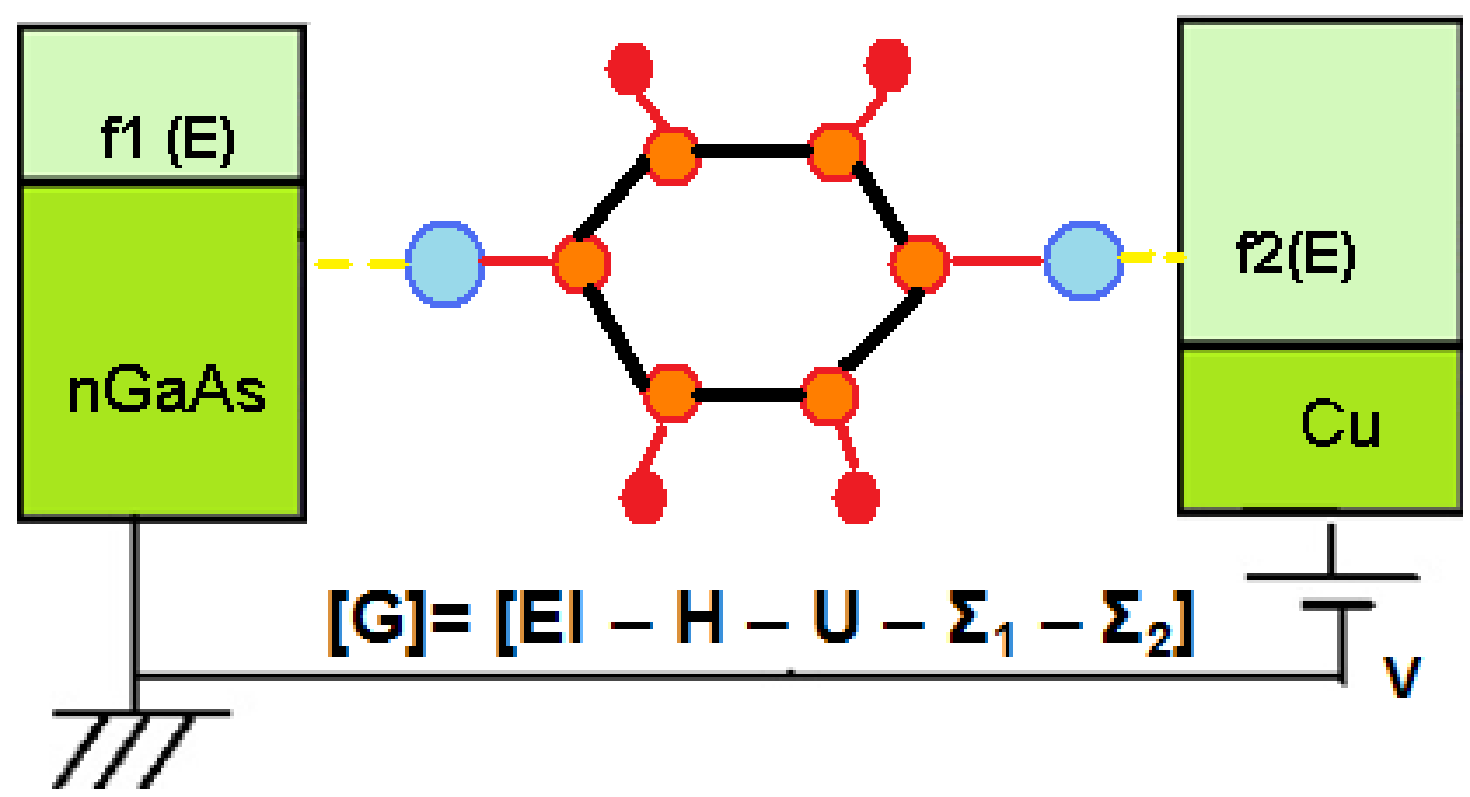

**Figure 4b.** Simplified model of Device.

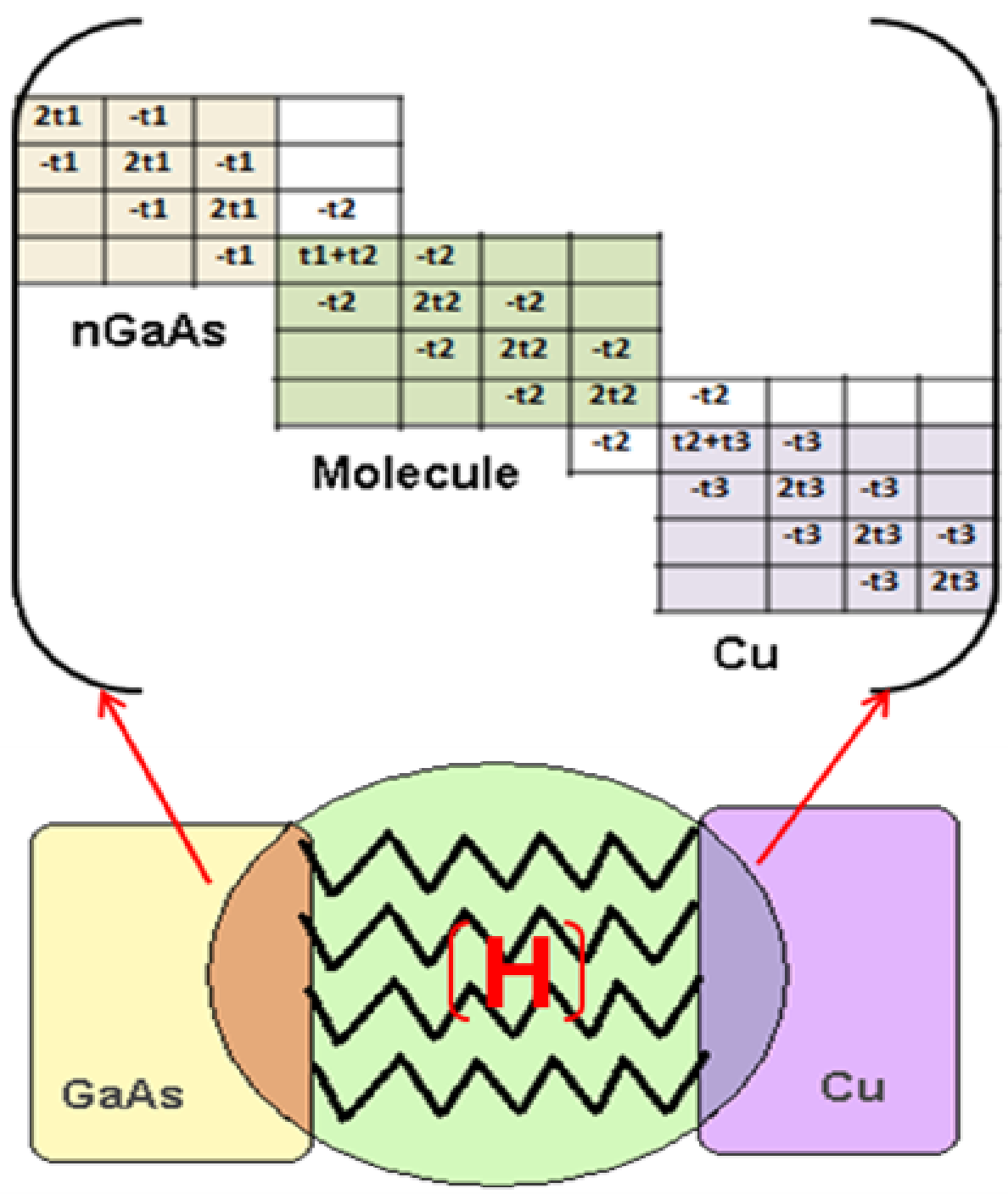

**Figure 4c.** Hamiltonian representing the potentials along the grid points

Model J-Vs based on the NEGF formalism were compared with experimental J-V characteristics in Figure 1 and we explain herein the methodology followed to lead to these fits. Drawing from our model results, we attribute the distinct slopes in forward bias to two concurrently occurring conduction channels: one dominated by thermionic emission over the semiconducting Schottky barrier at reverse bias/low forward bias, and the other by tunneling through the molecular barrier at higher forward bias. The former was captured by the Fermi functions (F), while the latter was captured by the quantum mechanical transmission T (see Eq. [8] of Methods section). It is worth emphasizing at this stage that the two processes are not strictly separable into a simple product of quantum mechanical and thermal probabilities. Instead, Eq. [8] can be shown to involve a *convolution* between temperature-independent quantum tunneling T and a thermal broadening function proportional to dF/dE. [10]

In reverse bias, current is dominated by electron flow from metal to semiconductor, which sees the Schottky barrier due to metal-semiconductor work function difference as well as the tunnel barrier through the molecule. However since the tunnel barrier is orders of magnitude smaller than the Schottky barrier, we attribute the slope and magnitude of the J-Vs in reverse bias to Thermionic emission. At low forward bias however, tunnel transport becomes significant due to the decreasing Schottky barrier. At a critical applied voltage in the forward bias, the decreasing and thinning Schottky barrier becomes transparent to electrons, and tunneling through the molecule becomes the dominant transport mechanism. The transition from the Schottky mediated to the tunneling mediated transport as observed from our NEGF calculations occurs when the GaAs depletion width ($W_d$) was equal to the Debye length ($L_d$) and its height equals the thermal energy.

**Molecule as a voltage divider**

Organic molecules can influence current flow through semiconductors in a variety of ways: by providing tunable resonant tunneling states or gate-able trap states that can lead to hysteresis and low-frequency noise, by inducing dipoles that influence the band-bending, as well as through vibronic and spintronic scattering channels for forward or backward scattering. Explaining such complex phenomena demand sophisticated modeling efforts incorporating the molecular chemistry and band-structures. For a molecule with a wide HOMO-LUMO gap like the alkane series however, the physics lies primarily in the electrostatics, namely, in its interfacial charges and barriers.

Figure 5a shows the schematic potential profile across the MMS junction under reverse bias. At equilibrium (no bias), there is a built-in voltage $V_{bi}$ across the device junction arising from the work function difference between the metal and semiconductor. Upon application of a voltage bias, a fraction $V_1$ drops across the semiconductor depletion region ($W_d$) representing the Schottky barrier, while $V_2$ across the length of the molecule ($L_m$) increasing the tunnel barrier (shown as blue lines in Figure 5a). The potential due to the Schottky barrier is assumed to vary quadratically across $W_d$ and that due to the tunnel barrier is varied linearly across $L_m$ in our model.

In the absence of any interface charge regions (dipoles), the voltage division is solely determined by the relative dielectric constants and lengths of the semiconductor depletion width as well as molecule.

$$(\varepsilon_{GaAs} V_1)/(W_d/2) = (\varepsilon_{Mol} V_2)/Lm \qquad \textbf{[1a]}$$

$$V_1 + V_2 + V_{bi} = V_{app} \qquad \textbf{[1b]}$$

where $L_m$ is the molecular length, $Wd = \sqrt{2\varepsilon_0 \varepsilon_{GaAs} V_1 / e^2}$, $N_D$ is the doping-dependent GaAs depletion width, $N_D$ is the doping density, $\epsilon_{GaAs}$ and $\epsilon_{mol}$ are the dielectric constants of GaAs and molecule respectively and $\epsilon_0 = 8.854 \times 10^{-12}$ F/m is the permittivity of free space. The factor of 2 on the left of Eq. [1a] arises because the voltage drops quadratically rather than linearly in the depletion region. The intrinsic (gas-phase) molecule-dependent polarizability enters our analysis through $\epsilon_{mol}$ above.

To account for interfacial charges arising from SAM induced dipoles and charge-transfer dipoles at the molecule-GaAs interface, it is straightforward to include a third barrier, which leads to an additional capacitive drop in band-edge from the positive to the negative terminal (shown as the red lines of the Energy band diagram in Figure 5a). The transport model can then be used to study the dependence of the shape and magnitude of the transport (J-V) characteristics on variations of the Schottky barrier ($V_1$), the tunnel barrier ($V_2$) and the barrier due to interfacial charges ($V_{dip}$), through using a circuit model of the voltage divisions as shown in Figure 5b and Eq. [1b] now modifies to:

$$V_1 + V_2 + V_{bi} - V_{dip} = V_{app}$$

To better understand the role of the molecule as a voltage divider in our MMS junction, we varied the molecular properties- dielectric constant, length and induced dipoles as parameters in the NEGF model and studied the effect on the J-V characteristics.

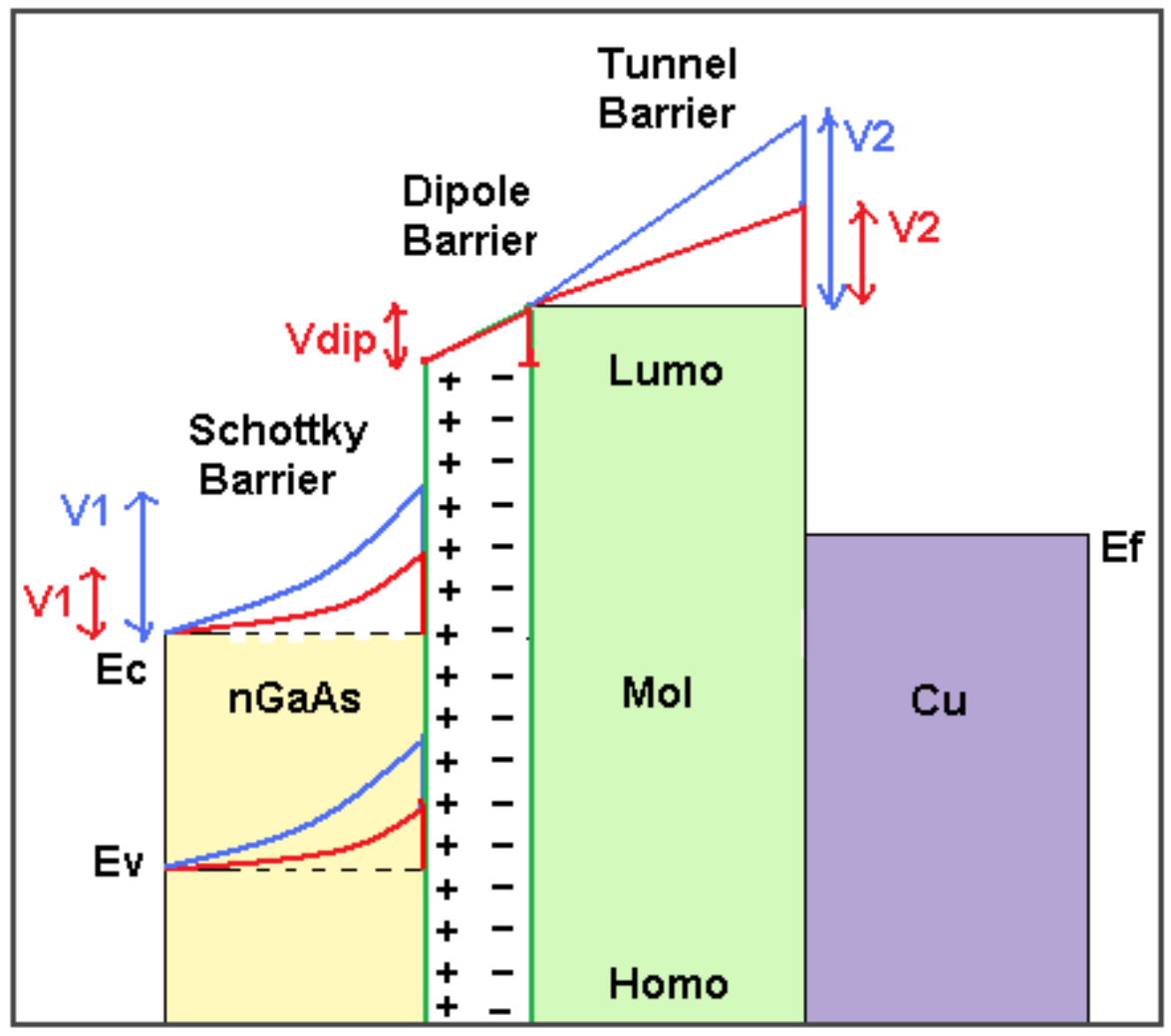

**Figure 5a.** Energy band diagram of the MMS device junction under reverse bias with $V_1$, $V_2$ and additional barrier $V_{dip}$ as the Schottky, Tunnel and Dipole barriers. Blue lines designate the band diagram including contributions just from V1 and V2, while the red lines show the effect of an additional Vdip on the voltage drops within the band diagram.

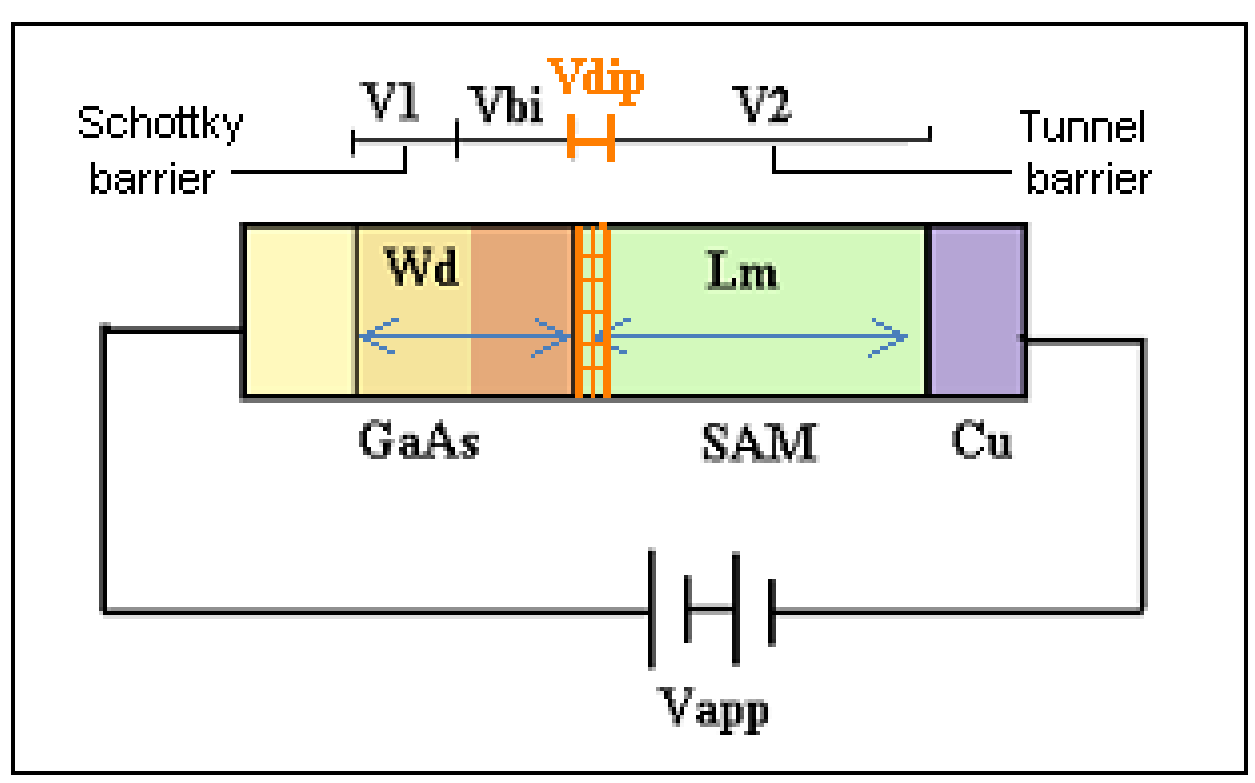

**Figure 5b.** Circuit diagram of device

**Dielectric constant of molecule: Effect on Schottky transport**

Upon varying the dielectric constant of the molecular layer ($\varepsilon_{mol}$), which effectively varies the ratio $V_1/V_2$ (Eq. [1]), the J-V characteristics were predominantly affected in the reverse and low forward bias (V < 0.2V) regions where Schottky barrier mediated thermionic transport was dominant, as shown in Figure 6a. This was also apparent from the increasing degree of asymmetry with increasing $\varepsilon_{mol}$, arising from lower current densities due to increasing of the Schottky barrier.

**Length of molecule: Effect on tunneling transport**

With increasing $L_m$, voltage drop across the molecular layer dominated over that due to the Schottky barrier (similar to our observations with decreasing $\varepsilon_{mol}$ in Figure 6a). Figure 6b shows the model J-V characteristics for variations to the tunnel barrier, obtained by varying the molecular length ($L_m$ varied for (CH2)n-COOH with n=7, 10 and 15). Upon increasing $L_m$, three significant features were observed in the model characteristics – (i) the kink due to transition from Schottky barrier to tunnel barrier mediated transport was shifted to earlier voltages; (ii) the characteristics were more symmetric; and (iii) the current magnitudes were exponentially reduced at high forward bias.

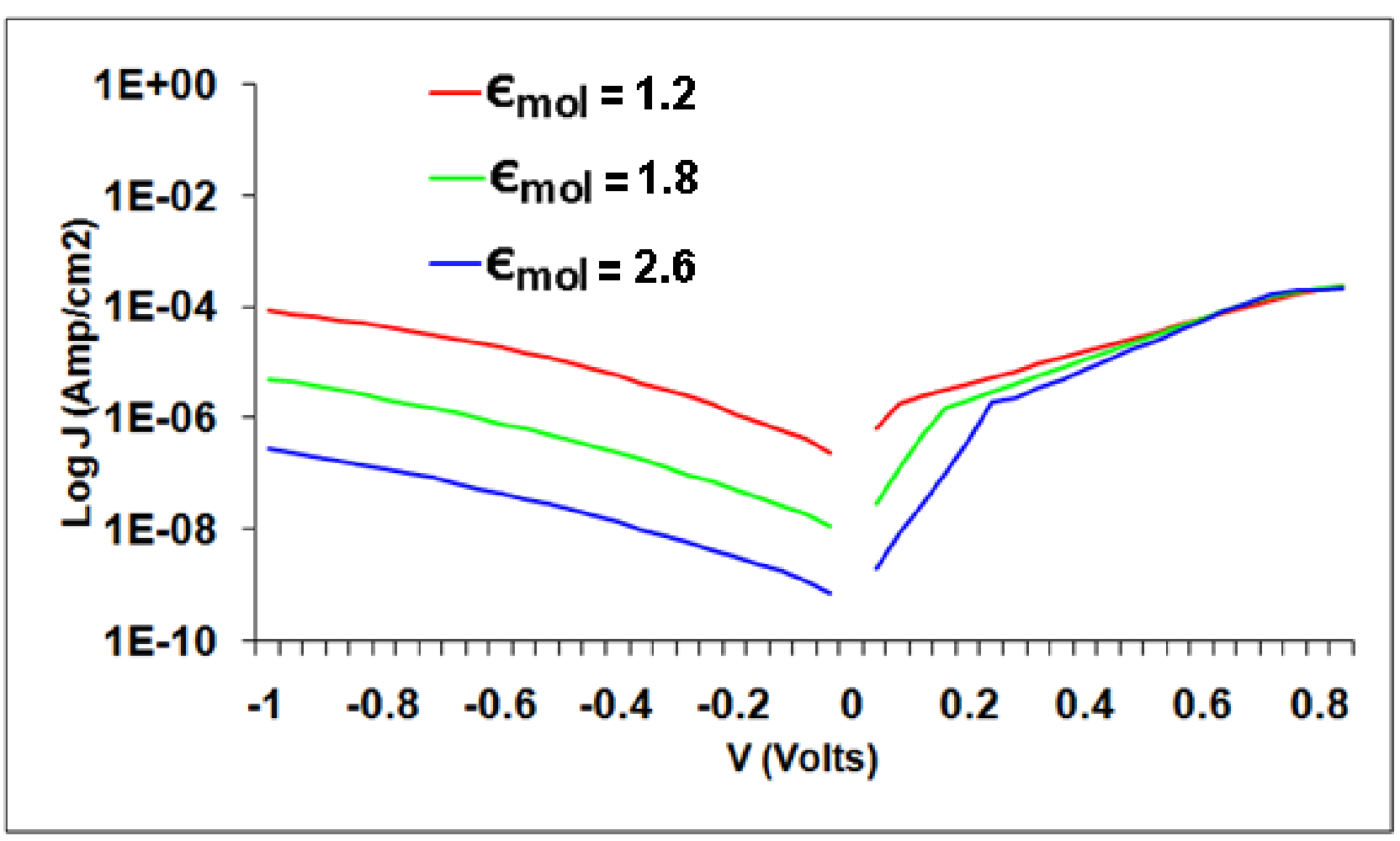

**Figure 6a.** Varying $\varepsilon_{mol}$ to modulate $V_1/V_2$ predominantly affects the reverse bias and low forward bias regimes of the computed J-V characteristics.

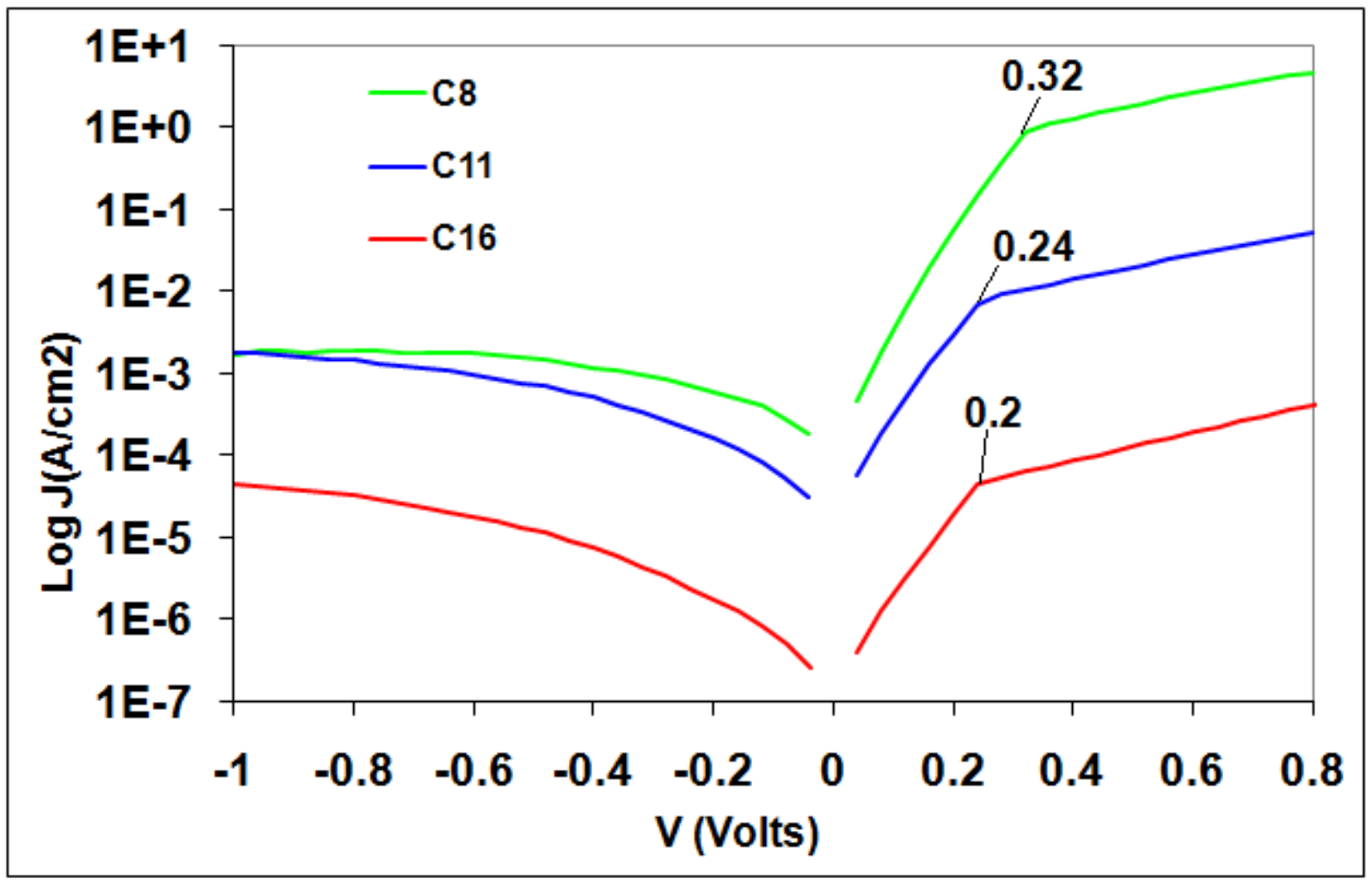

**Figure 6b.** Varying Tunnel barrier ($V_2$) or molecular length (L) affects only the high forward bias regimes of the computed J-V characteristics. Error bars indicate experimental data on C16.

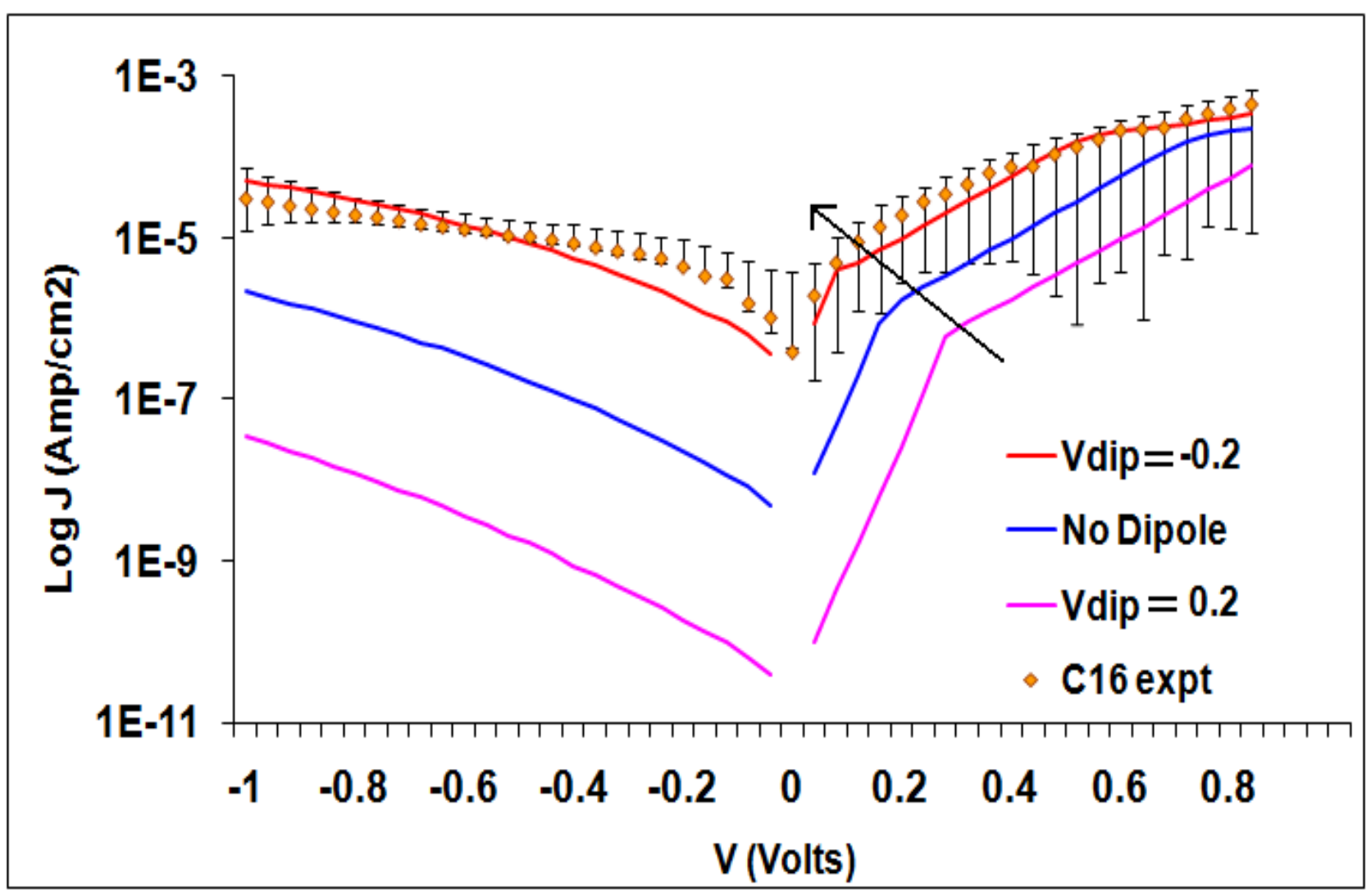

**Figure 6c.** A large negative dipole barrier is needed to shift the kinks at the crossover point to lower voltage values consistent with experiments

The asymmetric J-V characteristics arise from the Schottky barrier that promotes conduction in one direction and suppresses it in the other (analogous to a p-n junction). Increasing $L_m$ reduces this barrier, which in turn makes the J-V more symmetric. This results in an earlier onset voltage for tunnel transport explaining the shift in the kink. Finally, for high forward biases, where Schottky barrier becomes negligible and tunneling is the only transport mechanism, the currents drop exponentially and more significantly for longer length SAMs. It should be noted that current magnitudes also drop in the low forward and reverse bias regimes with increasing $L_m$, indicating that tunneling is still a contributing factor to transport even though the Schottky barrier might be playing a more significant role. This can be observed specifically in the reverse bias regime, where current magnitudes for C8 and C11 SAMs seem comparable, while those for C16 SAMs are much smaller. C16 being a much longer molecule has a higher tunnel barrier thus leading to much lower currents.

**Molecular dipoles: Need for an additional barrier**

While the qualitative trends can be adequately explained by including just $V_1$ and $V_2$ in the model, the extent of shift of the kink requires a larger voltage drop than our model captures. In our model, the voltage division was constrained by the overall applied voltage, the built-in potential, and the relative molecular layer and depletion layer capacitances. An additional shift in the kink voltage required an extra capacitive voltage drop, such as due to interfacial dipoles- $V_{dip}$. It would also renormalize the current levels, so that fitting the J- Vs would yield the capacitive drop across the interfacial dipole layer.

This can be observed in Figure 6c, where we vary the C16 dipole barriers from -0.2V to +0.2V. The model J-Vs fit the experimental characteristics only when a dipole barrier of -0.2V was invoked. Furthermore, it may be noted that relative to the "no dipole" ($V_{dip}$=0) curve, the characteristics based on negative $V_{dip}$ values (due to molecules with positive dipole moments) show kink shifts to earlier voltages and the characteristics based on positive $V_{dip}$ values (due to molecules with negative dipole moments) show kink shifts to later voltages. These shifts in the kink with varying molecular dipole moment and direction were also observed within prior experimental characteristics, [20, 21] but coupling to our quantum transport model enabled quantitative elucidation of the respective interfacial dipole barriers.

***Optimized fits to experiment***

Model and experimental J-V characteristics for each molecular junction (C8, C11 and C16 SAMs) were optimized for a good fit to experiment (model curves within the error bars of the experimental characteristics) by choosing appropriate values for SAM related parameters from references such as LUMO/HOMO levels, dielectric constants, molecular lengths, tilt angles and packing densities, while the interfacial charge induced dipole barrier was varied independently. As discussed in the previous section, attempts to fit without the dipole barrier showed a mismatch between model and experimental characteristics, particularly in the Schottky region. A final best fit as shown in Figure 1 was obtained after varying the Vdip barrier. The various parameter values used for the final fit are summarized in Table 1. The various optimized parameters used for fits are discussed in the next section.

**Table 1.** Model parameters used to fit experiments

| | **Lm[Å]** | **$t_2$[eV]** | **$t_3$ [eV]** | **Lumo [eV]** | **$\varepsilon_{MOL}$** | **Vdip [V]** |
|---|---|---|---|---|---|---|
| **C16** | **22** | **2.2** | **0.06** | **2.5** | **1.8** | **-0.2** |
| **C11** | **16** | **1.8** | **0.06** | **2.5** | **1.8** | **-0.18** |
| **C8** | **8** | **1.4** | **0.06** | **2.2** | **1.6** | **-0.04** |

## Model parameters

### *SAM Length*

Lengths of C16, C11 and C8 for the model were selected to be 22 Å, 16 Å and 8 Å, respectively, which correspond to values reported within prior work for corresponding alkane thiols with same number of carbon atoms (values experimentally measured in [7]). This length may also be theoretically derived from the formula [10]

$$t = 1.26*(n-1)*\cos\Phi + 1.85 + x \qquad \textbf{[2]}$$

where, x is the COO_Cu bond length, 1.26Å and 1.85Å are the C-C and C-S bond lengths, [13] respectively.

***Wavefunction overlap***

The wave-function overlap values for the semiconductor-molecule ($t_2$) and molecule-metal ($t_3$) junctions reflect qualitatively the coupling at the respective interfaces. As shown in Table 1, we used similar values for C11 and C16 SAMs, but lower values for C8 SAMs.

***LUMO level***

The Lowest Unoccupied Molecular Orbital (LUMO) level for C16 was chosen to be 2.5eV, based on the photoemission data from similar MMS junctions with SAMs of $C_{15}CH_2S$ [4] (conduction band edge of GaAs=1eV). We set the corresponding value to be the same for C11 SAMs. For C8 SAMs, we chose a lower LUMO level value of 2.2eV. This is attributed to Image Force Lowering effect which is significant only for the shortest molecule C8, not observed in C11 and C16. [19]

The maximum electric field across C8 for tunneling dominated transport was ~ $2.5*10^6$V/cm. It was calculated by dividing the maximum voltage drop $V_2$ across the molecules (which is ~0.2V at $V_{app}$=1V) by the molecular length (8Å). The maximum Image Barrier lowering ~ 0.5eV was then calculated by: [19]

$$\Delta\Phi_m = [q*|E\max|/4\Pi\varepsilon_{mol}\varepsilon_0]^{1/2} \qquad \textbf{[3]}$$

where $\Delta\Phi_m$ represents the reduction in barrier. However since the Electric field varies with distance (inside the molecule) as well as with applied voltage, the actual barrier lowering would be less than 0.5eV. By varying the LUMO level between 2.5eV and 2eV, we obtained the best fit with experiment for a reduced barrier of 0.3eV (LUMO=2.2eV).

***Dielectric constant***

We used dielectric constants of ~1.8, for C16 and C11 SAMs, which are close to the values used literature for alkane thiol SAMs. [22] For C8 SAMs we used a lower dielectric constant of 1.6, to account for the lower coverage for shorter SAMs. [21, 23]

***Packing density***

A packing density of ~$5*10^{13}$ $cm^{-2}$ was used for C16 and C11 SAMs. Due to more repulsive intermolecular interactions within shorter SAMs, [21] longer SAMs are more stable and able to form at higher packing densities. Since C8 SAMs are shorter, a packing density of $10^{13}cm^{-2}$ (less by a factor of ~5 than the higher length SAMs) was used.

***SAM Tilt***

The tilt angle for C16 and C11 alkyl chain SAMs to the substrate normal was assumed to be $30^{\circ}$, while the tilt angle was assumed to be to 45º for C8 SAMs.

To compare the dipole values determined from our NEGF fit to the transport data with expected dipole moments of the molecular layers, we did Quantum chemical *'ab-initio'* calculations for the dipole of a free molecule and and of that in a chemisorbed system. These results are detailed in Table 2.

**Table 2.** Quantum chemical calculations for dipole moments $\mu_{Mol}$ -isolated gas phase molecules and $\mu_{CLUSTER}$ -chemisorbed cluster systems.

| NEGF Fit to Experimental data | | | | Quantum Chemical Calculations | |
|---|---|---|---|---|---|
| $V_{dip} = N_{mol}\,\mu\,\cos\theta/\varepsilon\,\varepsilon_{mol}$ | | | | Gas phase | Cluster |
| | $N_{MOL}$ [cm$^{-2}$] | $\Theta$ [Deg] | $\mu$ [De] | $\mu_{MOL}$ [De] | $\mu_{CLUSTER}$ [De] |
| C16 | ~5e10$^{13}$ | 30° | 2.2 | 3.1 | 2.19 |
| C11 | ~5e10$^{13}$ | 30° | 1.98 | 2.49 | 1.76 |
| C8 | ~1e10$^{13}$ | 45° | 3.36 | 4.82 | 3.40 |

**Molecule induced Dipoles**

***Quantifying molecule induced dipole barriers***

After studying the effect of each parameter ($\epsilon_{mol}$, $L_m$ and $V_{dip}$) on voltage division and the resulting J-V characteristics, we obtained $V_{dip}$ barrier values of -0.2, -0.18 and -0.04 V for C16, C11 and C8 SAMs, respectively. We elucidate herein the physical significance of the fitted $V_{dip}$ barrier values. From dipole-induced barrier values $V_{dip}$ we computed the average dipole moment μ for the molecule in SAM from the relation [24]

$$V_{dip} = N_{mol}\mu\cos\theta / \varepsilon_0\varepsilon_{mol} \quad \textbf{[4]}$$

where $N_{mol}$ is the packing density of SAM, θ is the tilt angle of the SAM on substrate and $\varepsilon_{mol}$ is the dielectric constant of the SAM. We also performed Quantum chemical calculations to obtain dipole values of these molecules in gas-phase ($\mu_{MOL}$) as well as in the chemisorbed system ($\mu_{CLUSTER}$). We observed that the dipole moment (μ) determined from the model fits matched the values obtained from cluster model calculations, while this value was lower than the gas-phase dipole moment ($\mu_{MOL}$) by a factor of ~0.7, as shown in Table 2. We attribute this dipole value reduction to the depolarization caused during SAM formation on semiconducting surfaces, which results in a net surface dipole. [25] The value of this surface dipole depends not only upon the magnitude of the gas phase molecular dipole and the dipole due to the chemical bond between the isolated molecule and the substrate surface, but also on a number of other factors, such as i) charge re-distribution in SAM ii) intra-molecular structural rearrangement and iii) charge transfer from SAM to/from substrate which come into play when an aggregate of molecules forms on a surface. [25, 26, 27] These effects play a role in reducing the overall molecular dipole moment of the SAM layer.

Quantum chemical calculations using Density Functional Theory (DFT) were used to obtain the gas phase molecular dipole[28] values for C16, C11 and C8. The results (summarized in Table 2) for these isolated molecules were in agreement with literature. [29] Quantum chemical cluster- calculations were then performed to estimate molecular dipoles in chemisorbed systems - molecular aggregates on GaAs substrate surface. The density of molecules used in the cluster (see Methods section) for C8, C11 and C16 were similar to the ones used in NEGF modeling as can be seen in Table 2. This confirms that the NEGF computed dipole moments obtained from $V_{dip}$ barriers required to fit the experiment transport data for molecular junctions compared well to the dipole moments obtained from the cluster calculations.

***Modulation of semiconductor band bending: Interfacial charges***

Dipoles from a molecular monolayer can induce charge at the molecule-semiconductor interface due to the following reasons: [25, 26, 27] i) contribution from the perpendicular component of its dipole moment that depends on the dipole moment of gas phase molecule and molecular tilt within the SAM, ii) charge transfer to or from the SAM which is dependent upon the electro-negativity of functional groups within the SAM, packing density of the SAM and the alignment of molecular HOMO and LUMO levels to semiconductor bands; and iii) depolarization effects due to intermolecular interactions in a SAM. Our model J-V characteristics demonstrate that interfacial charges due to SAM-induced dipoles affect the voltage division arising from the Schottky and tunnel barriers, and hence the net transport through the junction. These SAM-induced charges at the molecule-semiconductor interface influence the Schottky barrier by affecting the net band-bending at the semiconductor surface.

To isolate the effect of dipole barrier from the Schottky barrier and understand how its magnitude and sign affects the direction of band-bending, we plotted the band-bending (green, blue and red lines in Figure 7) for a molecular layer of length equivalent to the C16 SAM, but with dipole barriers varying as: $V_{dip}$=-0.2V, -0.04V and +0.2V, without including $V_{bi}$ (built-in voltage attributed to semiconductor-metal work function differences). We observed that the band-bending can be switched from downward to the upward direction due to the sign change of the SAM-induced barrier ($V_{dip}$) causing a modulation of the surface charge (positive to negative semiconductor surface charge).

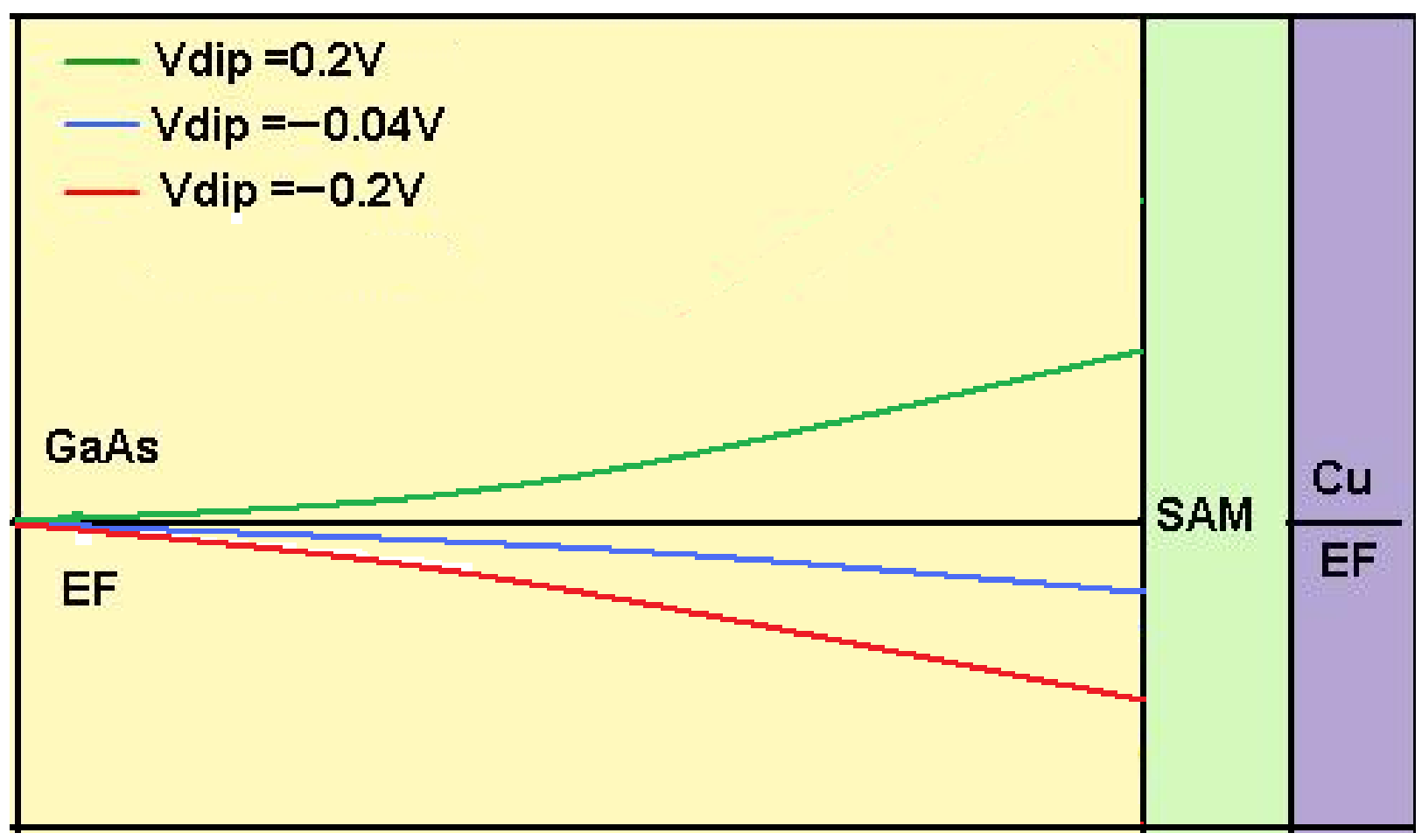

**Fig 7.** Calculated band bending in nGaAs induced as a result of varying dipole barriers from functional groups in SAMs of length equivalent to C16 SAMs. The band-bending depicted does not include built-in work-function $V_{bi}$. NEGF fitted dipole barrier values Vdip = 0.2V, -0.04V, -0.2V (Green, blue and red lines) are shown. (Band bending not drawn to scale, x-axis shows normalized molecular length)

From our NEGF model results, the fitted $V_{dip}$ barrier value increased in the negative direction for longer molecules ($V_{dip}$=-0.2 for C16) versus shorter molecules ($V_{dip}$=-0.04 for C8). Based on these observations, we can conclude that a more downward molecular-dipole induced band bending occurs for C16 (and C11) SAMs versus C8 SAMs. Furthermore, gas-phase Quantum Chemical calculations confirm a positive dipole (inducing positive surface charge at the surface) for all the three molecules studied herein. From this we can establish conclusively that a) SAMs of C8, C11 and C16 induce a positive surface charge at the nGaAs surface (due to the downward band-bending calculated for the fitted $V_{dip}$ barrier value) and b) longer SAMs studied herein (C16 and C11) induce a higher surface charge (as elucidated from the higher fitted $V_{dip}$ barrier values). The positive surface charge arises from the positive dipole in Table 2, thus generating a downward band-bending. This result, explained quantitatively herein for our experimental transport data on molecular junctions on GaAs agrees with qualitative observations in the literature of shifts in the kink of the transport data with interfacial charges, as reported for CN terminal alkyl SAMs on nGaAs. [30, 31, 31]

The dipolar results in Table 2 depict that in comparison to C11 and C16, the gas phase dipole moment of C8 molecule is higher. On one hand charge transfer should be easier on C8 SAMs versus the longer SAMs, due to higher depolarization effects on shorter SAMs such as C8, as well as due to the lower LUMO level required to fit our experimental transport data for C8 versus C11 and C16, based on the prominent role of image force effects. However, on the other hand, these effects need to be balanced against the lower packing density (by a factor of 5) and higher degree of tilt for C8 SAMs compared to C11 and C16 (which have similar dipoles and packing). Based on our results, we observe that the lowered C8 packing and higher tilt, in fact, dominate over the expected higher depolarization effects on C8 SAMs.

## Conclusions

We presented a methodology that coupled experimental J-V characteristics of molecular junctions on n-GaAs to an NEGF based quantum transport model to quantitatively elucidate how the presence of a molecular layer (dielectric constant, length and induced dipoles) affects transport through the Cu-COOH$(CH_2)_n$SH-nGaAs (n = 7, 10 and 15) MMS junctions. The NEGF model was able to adequately describe the separable effects of Schottky, Tunneling and SAM-induced dipole barriers on transport. Variations to these barriers resulted in shape changes of the transport characteristics observed as changes in current magnitude, symmetry and cross-over voltages. Specifically the model was able to quantify a potential barrier value ($V_{dip}$) unique to each molecular SAM (C8, C11 and C16) and characterize its effect on semiconductor band-bending.

The dipole moments extracted from the experimental fits matched those obtained from *ab-initio* Quantum chemical cluster-based calculations for chemisorbed systems. These dipole moments were however lower than those obtained from *ab-initio* Quantum chemical calculations for gas phase molecules, the difference being attributed to depolarization and image effects that come into play during SAM formation and adsorption on a semiconductor substrate. Based on the fitted SAM-induced dipole barrier values we also conclude that $COO^-$ charges at the tail of longer alkanethiol molecules induce a greater degree of depolarization of the semiconductor surface compared to the shorter molecules. This higher depolarization can be explained in terms of the higher packing density of the longer molecular layers, as well as the greater voltage drop across larger separations for a capacitor plate model for the charges.

In future studies we aim to understand the effects of varying dopant type and concentration on the transport characteristics and semiconductor band-bending, including effects like Fermi Level pinning on the semiconductor surface. We also hope to conduct experiments on molecules with different HOMO-LUMO gaps, head-groups and tail-groups. Finally, we plan to extract the dipolar effects of the chemisorbed SAMs more accurately by including the metallic Cu layer in the cluster-based calculations, as well as using a periodic slab model instead of a cluster model.

## METHODS

Experiment In prior work we measured the experimental J-V characteristics [11] of an MMS junction with metal (Cu) – molecule (SH-(CH2)n-COOH) – semiconductor (n-GaAs) device junctions of varying device lengths (Figure 1). COOH terminal SAMs of C8, C11 and C16 were deposited on the patterned (7 µm X 7µm device area) n-type GaAs surface. Selective complexation and electro-less deposition of Cu was then carried out to form top contacts for transport measurements on multiple device junctions per SAM type.

**Richardson-Simmons Model** The Richardson's equation used for Thermionic emission, [19] modified for additional tunneling through the molecule was as follows:

$$J = A^{**}T^2[\exp(qV/kT)-1]\exp(-q\Phi_B/kT)\exp(-\beta L_{mol}) \qquad \textbf{[5]}$$

where, $A^{**}$ is the Richardson constant (4.4*120A/cm$^2$-K$^2$, taking into account back-scattering and quantum mechanical reflection). In the equation, T is the temperature (300K), V is the applied voltage, $\Phi_B$ is the zero bias built-in potential difference at the metal-semiconductor interface, ß is a tunneling attenuation factor, $L_{mol}$ is the molecular length. The value for ß was calculated from the following equation:

$$\beta = 2\sqrt{(2\Pi m\Phi_T / h^2} \qquad \textbf{[6]}$$

where $\Phi_T$ is the effective Tunnel barrier height through the molecule ($E_{LUMO}$ – eV/2, difference between LUMO level of molecule and the average potential of the electron at applied bias V).

**NEGF transport formalism** The longitudinal modes of the MMS device junction were described as a 1-D chain of grid points aligned along the transport direction (Figure 2a and 2b), so that the tri-diagonal device Hamiltonian spans the GaAs depletion layer, the molecule and a few atomic layers of the metal top contact (Figure 2c). The rest of the bulk GaAs and the metal Hamiltonian blocks were incorporated into a recursive algorithm to compute the contact surface Green's functions that determine their self-energies. The local hopping terms were averaged to create the corresponding on-site energies, amounting in effect to the matching of wave-functions and current densities (velocities) at each interface. The Hamiltonian matrix H was augmented with a matrix U to represent the potential profile at each respective grid point, as well as the wave-function overlap between adjacent grid points.

The Hamiltonian and self-energies were then incorporated into an NEGF model for current flow. Although the grid points for the model included only a 1D chain, full 3D effects and transverse modes were captured within the 2-D Fermi function F given by summing over transverse modes [10]

$$F(E)_{kxky} = \sum f_0(E + \mathrm{h}^2/2\mathrm{m_c}(\mathrm{k_x}^2 + \mathrm{k_y}^2)) = (\mathrm{m_c}\mathrm{S})\Phi \mathrm{h}^2 \ln/[1 + (\mathrm{E} - \mathrm{E}_F)/K_\mathrm{B}\mathrm{T}] \qquad \textbf{[7]}$$

where, $f_0$ is the regular 3-D Fermi function and S is the device cross-sectional area.

A critical determinant of the transport physics is the voltage-division across the heterojunction. The potential matrix U, used to augment the Hamiltonian, was defined to account for any electrostatic variations due to applied bias, matching the normal components of the electric displacement vectors at all charge-free interfaces. Charge injection/removal and level broadening were captured by the self-energy matrices $\sum_1$ and $\sum_2$. In principle, a lot of physics could reside in these self-matrices, capturing resonant tunneling diode behavior due to level slippage past the band-edges [16], as well as chemistry due to surface reconstruction [17]. In the interest of simplicity, and observing that the alkane levels sit too far from the band-edges to probe their detailed dispersions, we ignored the energy-dependence of these self-energy matrices.

The NEGF equations for transport were then formalized by first determining the Green's function

$$G = [EI - H - U - \Sigma_1 - \Sigma_2]^{-1} \qquad \textbf{[8a]}$$

The Transmission was calculated using the Fisher-Lee formula: [10 14]

$$T = Trace[T_1 * G * T_2 * G+] \qquad \textbf{[8b]}$$

Transport characteristics (J-V) were then generated by calculating current at each applied bias within the coherent NEGF (Landauer) formula, as given by:

$$I = 2e/h\int dET(E)[F1(E) - F2(E)] \quad \textbf{[8c]}$$

where, $F_1(E)$ and $F_2(E)$ are the aforementioned 2D fermi-functions keeping the bulk semiconductor and metal contacts in equilibrium. Since the F functions are proportional to the cross-sectional area S (Eq. [7]), dividing throughout by S yields the areal current density J. The results of the NEGF model are discussed in section 5.

**Ab-initio dipole calculations** Quantum chemical calculations using DFT to obtain the gas phase molecular dipole values for C16, C11 and C8 were implemented using the Gaussian 98 package [28]. Geometry optimizations were performed using both Hartree Fock (HF) and B3LYP functionals with a 6-31G* basis set for all atoms. Quantum chemical cluster-based calculations to estimate molecular dipoles in chemisorbed systems molecular aggregates on GaAs substrate surface were also performed using the Gaussian 98 package. The cluster size used included 4 molecules per cluster, since any further increase of size did not lead to significant changes in the dipole values. The density of molecules used in the cluster was $4e10^{13}$mol/cm$^3$ for C16 and C11 and $1e10^{13}$ for C8.

ACKNOWLEDGMENT Support for this work was obtained from NSF ECCS Award 0701505.